\documentclass{PoS}

\PoS{PoS(HEP2005)284}

\title {New results on  $e^+e^-\to hadrons$ exclusive cross sections         
from experiments with SND detector at VEPP-2M $e^+e^-$ collider 
in the energy  range  $\sqrt[]{s}=0.4\div 1.4$ GeV}

\ShortTitle{Results from SND}

\author{\speaker{Sergey Serednyakov}\thanks{seredn@inp.nsk.su}\\
        Budker Institute of Nuclear Physics, Russia}
\author {M.N.Achasov,  K.I.Beloborodov, A.G.Bogdanchikov,
A.V.Bozhenok, A.D.Bukin, D.A.Bukin, T.V.Dimova, V.P.Druzhinin,
V.B.Golubev, A.A.Korol, S.V.Koshuba, E.V.Pakhtusova, Yu.M.Shatunov, 
V.A.Sidorov, Z.K.Silagadze, A.N.Skrinsky, 
Yu.A.Tikhonov, A.V.Vasiljev,        
\\Budker Institute of Nuclear Physics, 
Novosibirsk State University, Russia}

\abstract
{ New results of the  $e^+e^-\to\pi^+\pi^-,         
~K^+K^-,~K_SK_L,~\pi^0\gamma,~\eta^0\gamma$ processes cross section
measurements are presented. The results are based on the 30 $pb^{-1}$ data, 
accumulated by SND detector at VEPP-2M $e^+e^-$ collider in the energy range 
$\sqrt{s}=0.4\div 1.4$ GeV during                     
1995-2000 years. The comparison with existing experimental data 
shows that the measurement accuracy is close to or better than  
the world average. For the $e^+e^-\to \pi^+\pi^-$ process the accuracy 
is $\sim 1\%$. This is important for calculation of hadronic contribution  
into the muon anomalous magnetic moment.  }

\FullConference{International Europhysics Conference on High Energy Physics\\
		 July 21st - 27th 2005\\
		 Lisboa, Portugal}

\begin{document}
\section{Introduction}        
Beginning from 1995,  experiments have been carried out at VEPP-2M $e^+e^-$
collider \cite{vepp2}  with SND detector \cite{sndnim}. 
The important part of the experimental program was  
study of the $e^+e^-$ annihilation into hadrons. The latest results of
this investigation  are presented in this talk \cite{prep05,pipi}.              
              
   VEPP-2M  $e^+e^-$ collider operated  during 1974-2000 years in the
center-of-mass (C.M.)               
energy range $E=0.4\div 1.4$ GeV. The luminosity depended on the energy, 
its average value was $L=2\times 10^{30}~cm^{-2}s^{-1}$ at $E\simeq 1$ GeV.        
In 2000 the experiments        
were completed and  the collider ring was dismounted. Now the new VEPP-2000        
$e^+e^-$ collider \cite{vepp2000} with higher maximum energy 2 GeV is under        
construction at the location of VEPP-2M.        
        
   The SND detector is general purpose nonmagnetic detector for low                  
energy $e^+e^-$ colliders. It consists of  the drift-chamber tracking        
system, three layer spherical electromagnetic calorimeter with        
1680 NaI(Tl) crystals, and  the muon detector with        
plastic scintillator counters and streamer tubes. The experiments at        
VEPP-2M were carried out in scanning mode of the collider energy range.        
The total accumulated luminosity is 30 inverse picobarns.

\section{The  $e^+e^-\to \pi^+\pi^-$ process}        
\begin{figure}
\includegraphics[width=.33\linewidth]{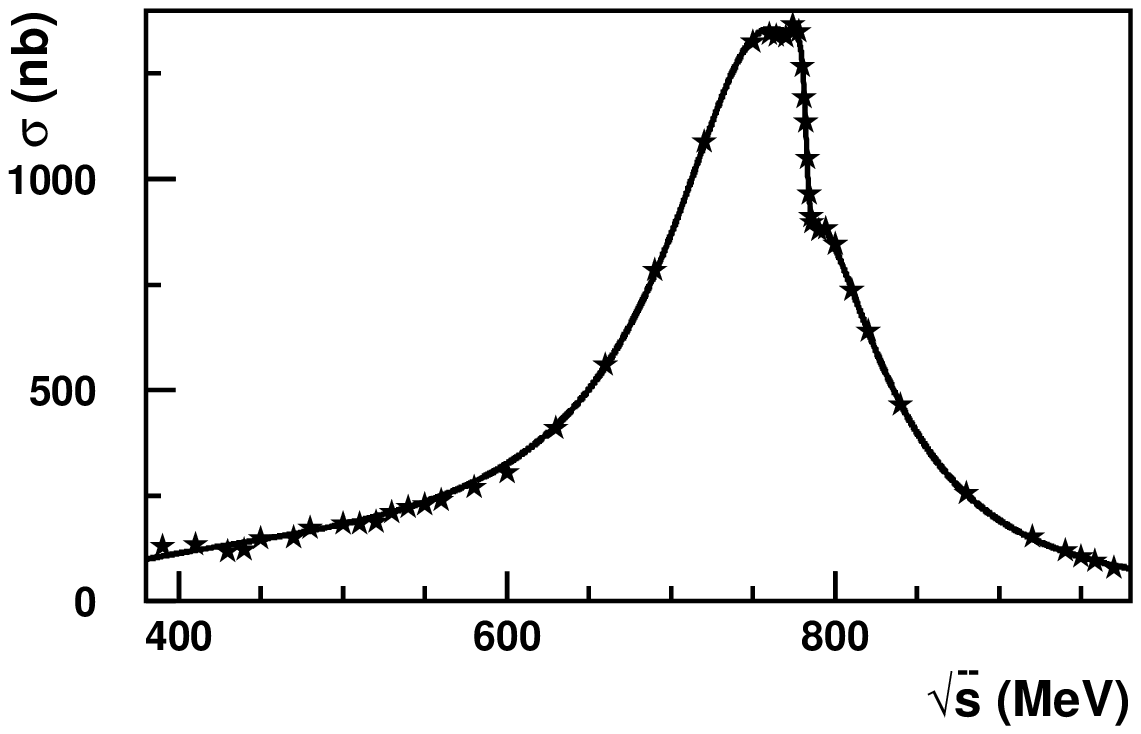}  
\includegraphics[width=.33\linewidth]{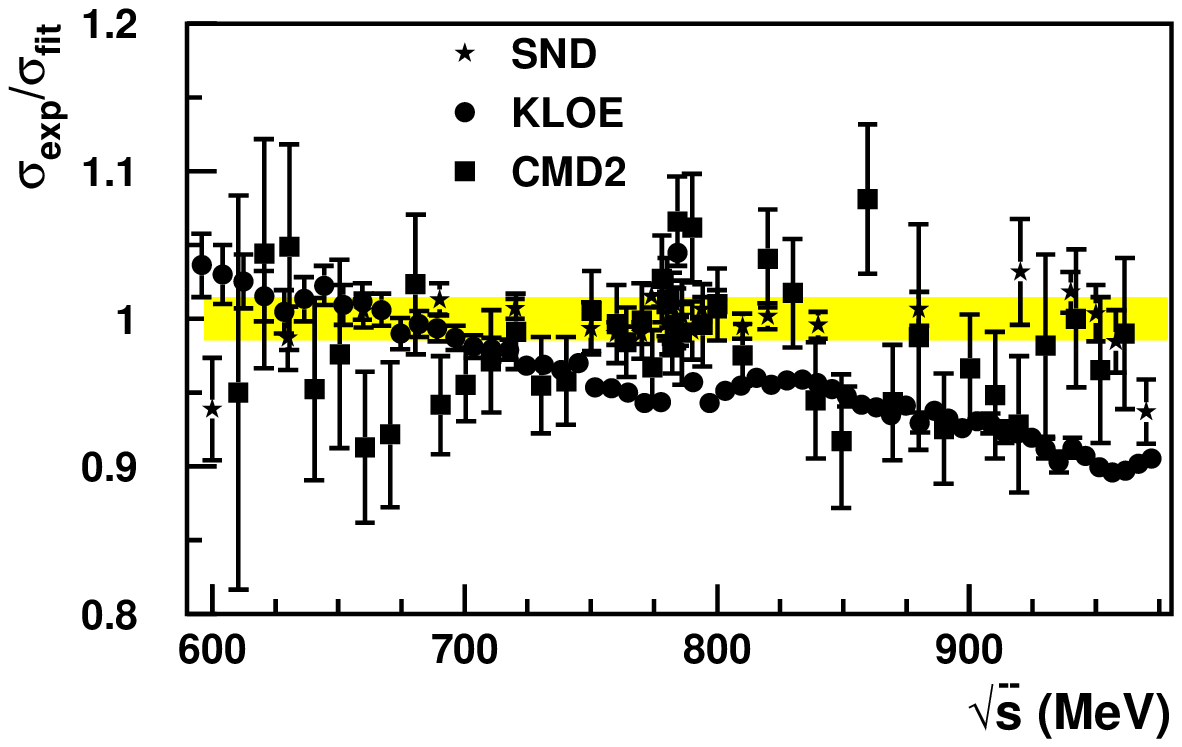}
\includegraphics[width=.33\linewidth]{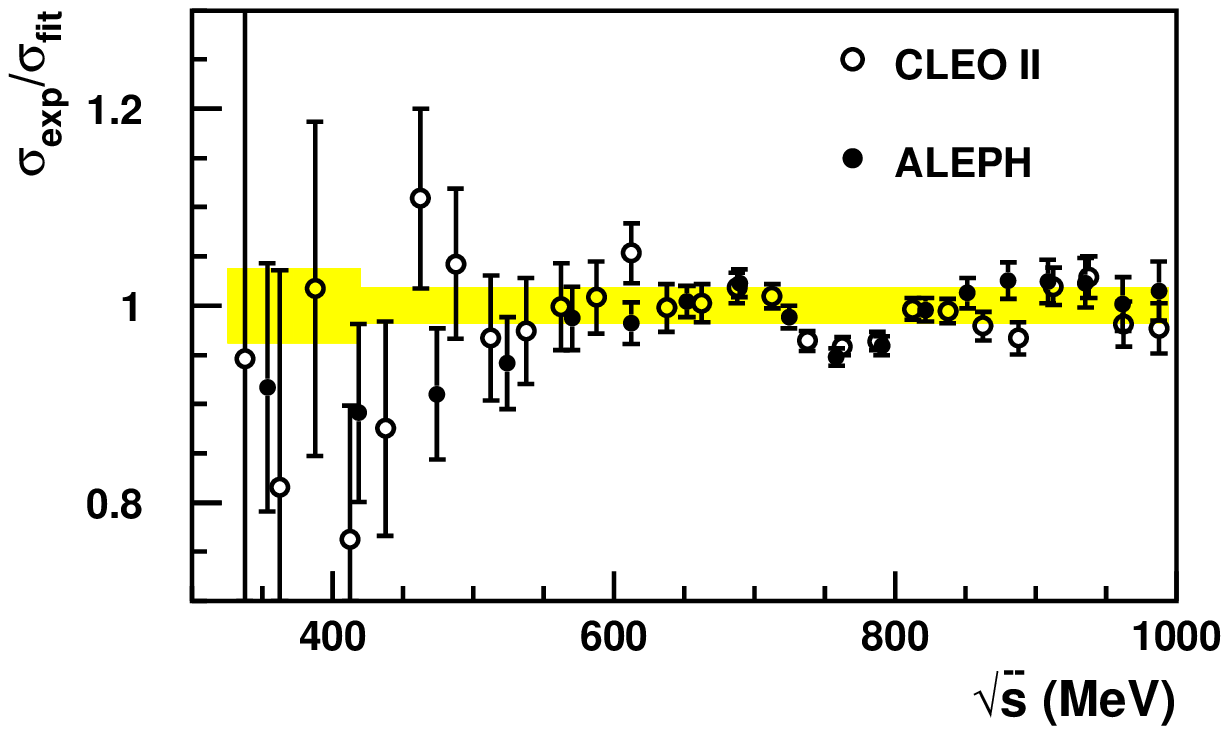}                
\caption{a - the measured cross section of the $e^+e^-\to\pi^+\pi^-$ process,     
the curve is the best fit; b - the ratio of the  $e^+e^-\to\pi^+\pi^-$ cross section  
obtained in CMD-2 \cite{kmd2} and KLOE \cite{kloe}  to the SND fit curve,  
the shaded area shows SND  systematics; c -  the ratio of the $e^+e^-\to\pi^+\pi^-$ 
cross section, calculated from  $\tau$-decay spectral function  
\cite{cleo2,aleph} to the isovector part 
of the SND cross section, the shaded area shows the joint        
systematic error. }        
\label{crspipi}
\end{figure}

  The study of the $e^+e^-\to \pi^+\pi^-$ process is important because        
this process  gives the major contribution into the total hadronic cross section          
at $E<1$ GeV and hence its contribution into hadronic part of muon anomalous  
magnetic moment (AMM)  $a^{hadr}_\mu$ is dominant $\sim 70\%$.          
In our work \cite{pipi} the  $e^+e^-\to \pi^+\pi^-$          
process was measured in the energy range $E=0.36\div 0.87$ GeV with 
integrated luminosity of 10 $pb^{-1}$. The number of  detected $\pi^+\pi^-$ events 
in the polar angle interval $55^o<\theta <125^o$ is $4.5\times 10^6$. 
The detection efficiency depends on energy and  
varies within the limits $(10\div 50)\%$. The main background 
comes from  the $e^+e^-\to e^+e^-$,  
$e^+e^-\to \mu^+\mu^-$,  $e^+e^-\to \pi^+\pi^-\pi^0$ processes and from  
cosmic rays. To select  the $\pi^+\pi^-$  events the neural network method was
used. The measured cross section is shown in Fig.\ref{crspipi}a.        
The systematic error of measurement $(1.3\div3.2)\%$ is determined mainly by        
the polar angle aperture error, uncertainties of the nuclear 
interactions of pions, pion identification and luminosity errors. 

The comparison of the measured  cross section with previous  measurements is shown in 
Figs.\ref{crspipi}b, \ref{crspipi}c. The reasonable agreement with
$\tau$-lepton data \cite{cleo2}, \cite{aleph}  and $e^+e^-$ measurement 
of CMD-2 \cite{kmd2} is seen. The KLOE data \cite{kloe} obtained by the 
radiative return method are in large disagreement with our results.
The calculated contribution into the muon AMM from  the $e^+e^-\to             
\pi^+\pi^-$ process is  $a^{hadr}_\mu~(\pi\pi) = (488.7\pm 7.1)~10^{-10}$   
with  relative  accuracy  1.5\%.  The $\rho$ and $\omega$  mesons parameters were measured 
in this work also  with accuracies better or comparable with the world average (see
\cite{pipi}).

  \section{The  $e^+e^-\to K^+K^-$ and  $e^+e^-\to K_SK_L$ processes}                  
    These processes are important because they give the considerable
contribution into the  total hadronic cross section at $E>1$ GeV and  into
the muon AMM. Besides this  cross sections depend on         
contributions from all light vector mesons $\rho ,~\omega ,~\phi$, their        
excitations $\rho\prime ,~\omega\prime ,~\phi\prime$ and possible unknown  
states. So measurements of  the cross section        
open the  possibility to study  all these states. Finally, the isovector        
part of the cross section can be used to test the  CVC hypothesis in $\tau$-decays.        
        
\begin{figure}  
\begin{minipage}{0.45\textwidth}
\includegraphics[width=.95\linewidth]{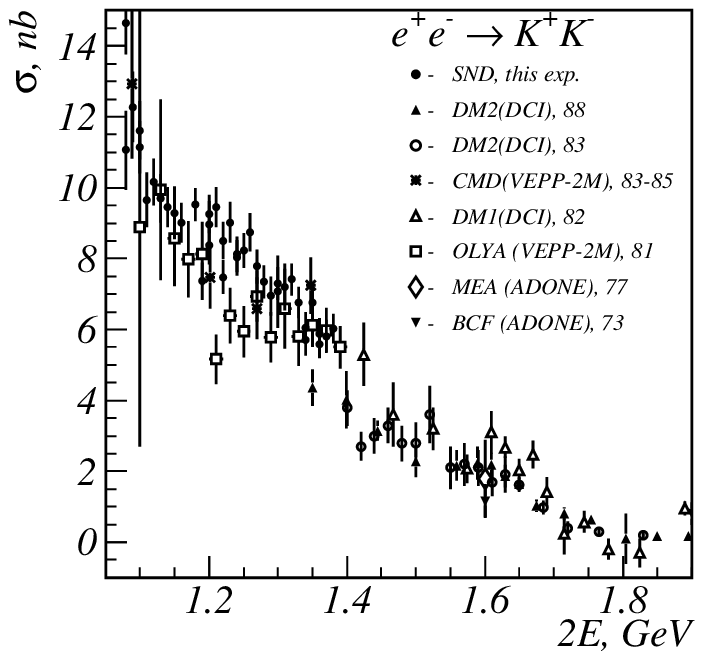}  
\caption{The measured cross section of  the $e^+e^-\to  K^+K^-$  
process }  
\label{kk}
\end{minipage}
\hfill
\begin{minipage}{0.45\textwidth}        
\includegraphics[width=.95\linewidth]{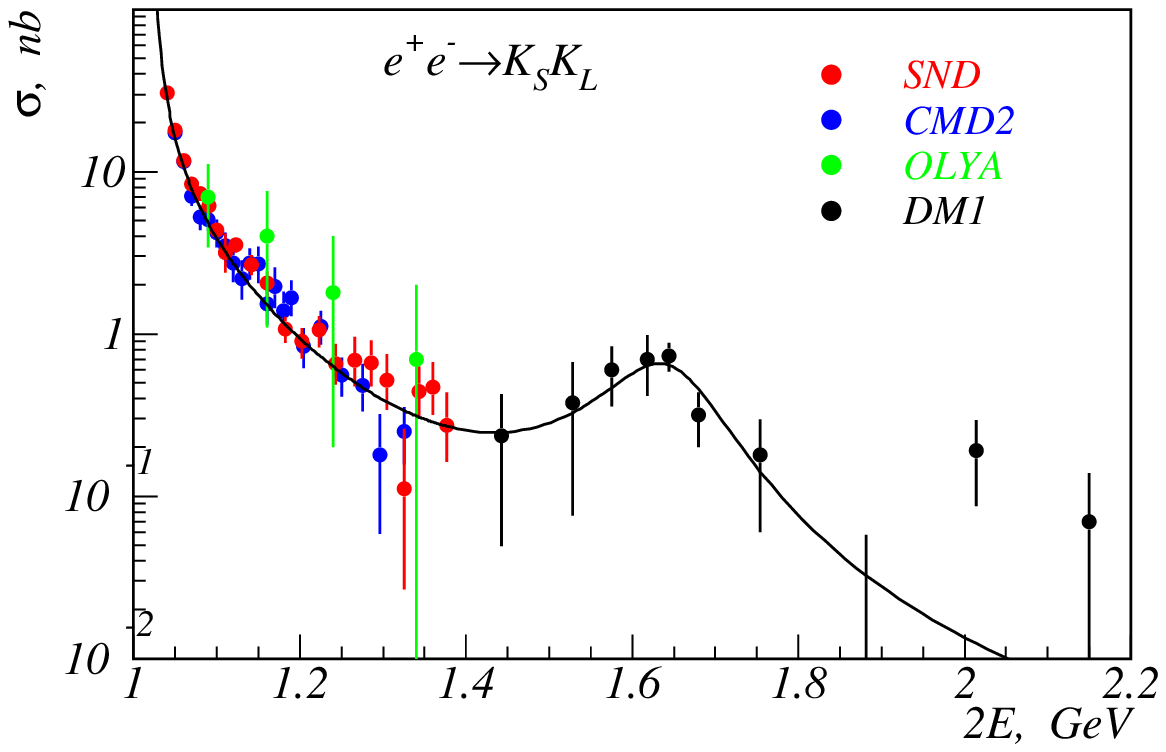}        
\caption{The measured cross section of  the $e^+e^-\to K_SK_L$ process,
the curve is the best fit to all data.}        
\label{kskl}  
\end{minipage}
\end{figure}

	     The kaon pairs production was studied in numerous                 
experiments, but the accuracy is still not sufficient  for full tests of the models        
mentioned above. In SND  experiment we measured the cross section        
in the energy range $2E=1.05\div 1.40 $ GeV with integrated luminosity        
$\Delta L\simeq 6~pb^{-1}$. To select  $K^+K^-$ events  the neural network   
analysis was used.  For  $K_SK_L$ channel investigation the neutral mode        
$K_S\to 2\pi^0$ was used. The measured cross section and results of
previous experiments  are shown in Figs.\ref{kk},\ref{kskl}.        
For $K^+K^-$ channel our cross section is the highest in comparison
with earlier data and confirms the exhibited  linear dependence versus energy. The
$\phi\prime (1680)$   resonance is not obviously seen. 
In $K_SK_L$ channel the SND result agrees with        
old data. The fit including all data, confirms $\phi\prime (1680)$.

		\section{The  $e^+e^-\to \pi^0\gamma$ and $e^+e^-\to
\eta^0\gamma$ processes}          
        
\begin{figure}  
\begin{minipage}{0.45\textwidth}
\includegraphics[width=.95\linewidth]{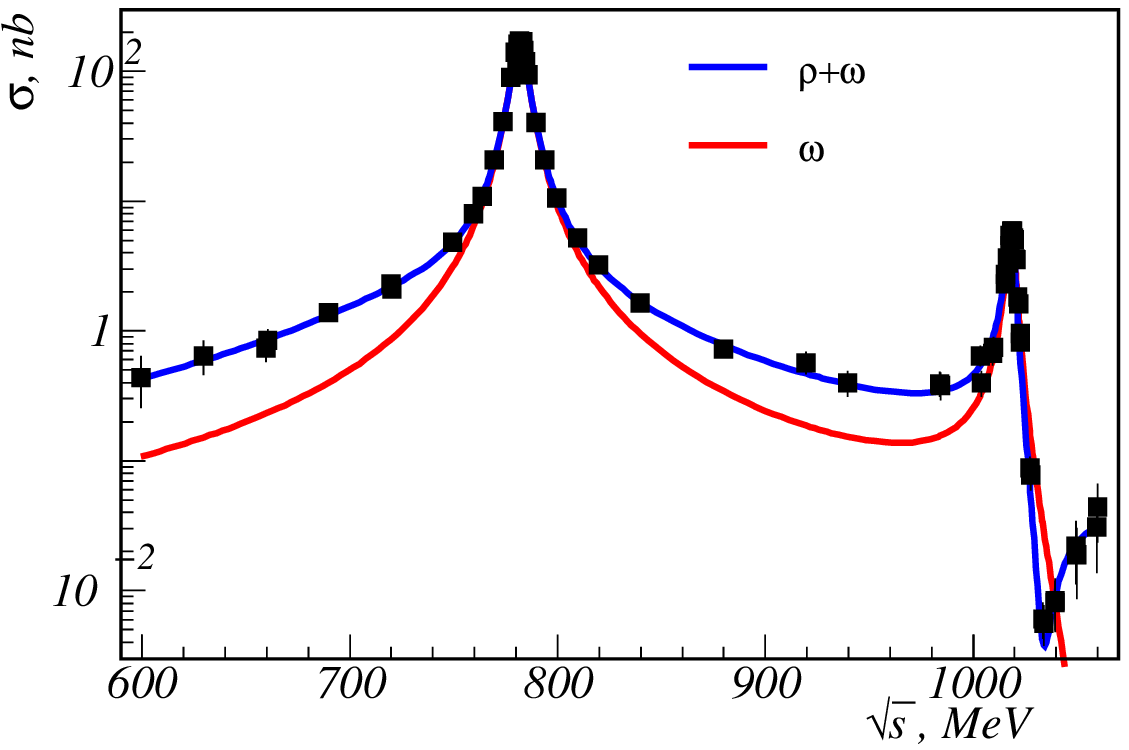}  
\caption{The measured cross section of  the $e^+e^-\to \pi^0\gamma$  
process, two fitting curves correspond to models with $\rho +\omega$ and $\omega$ 
intermediate states respectively. }  
\label{pigamma}  
\end{minipage}
\hfill
\begin{minipage}{0.45\textwidth}        
\includegraphics[width=.95\linewidth]{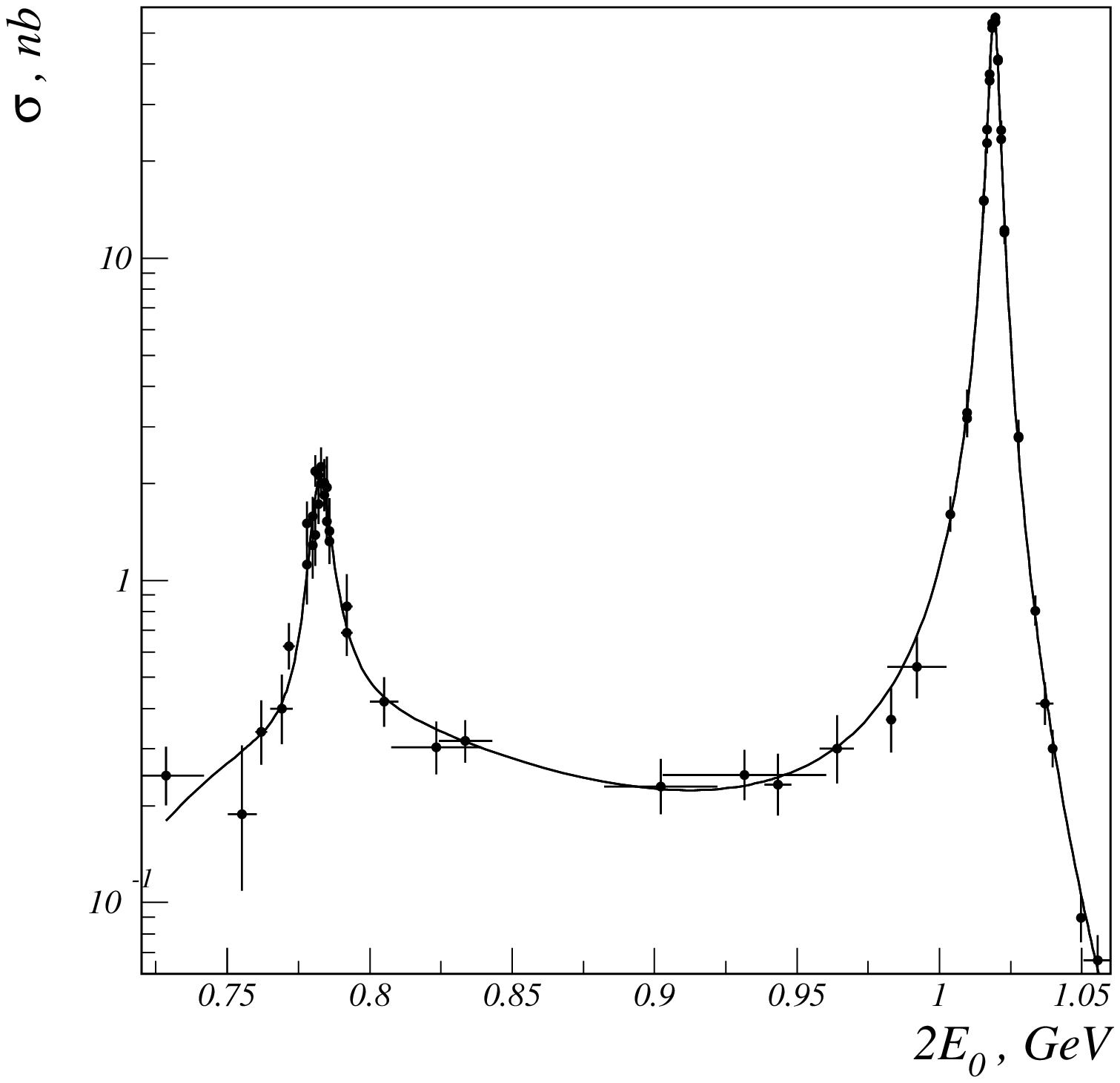}        
\caption{The measured cross section of  the $e^+e^-\to \eta^0\gamma$ process, 
the curve is the best fit.}     
\label{etgamma}
\end{minipage}
\end{figure}

    Both these reactions have significant cross sections in the
vicinity of   $\rho ,~\omega ,~\phi$. 
In SND experiment we measured the cross sections between        
resonances as well. For the first reaction the $3\gamma$ final 
state was used while for the second one we investigated two modes - 
with the $\eta\to 3\pi^0$ and        
$\eta\to \pi^+\pi^-\pi^0$ decays. The results are presented in Figs.        
\ref{pigamma},\ref{etgamma}, the curves being the vector meson dominance fits. The        
fitting parameters were vector meson decay parameters, which were measured   
in our work with accuracy comparable to or better than the world average. Below 
some prominent parameters measured in this work are listed:        
        
 $ Br(\rho\to\pi^0\gamma) = (5.02\pm 0.73)\cdot 10^{-4}$, 
  $ Br(\phi\to\pi^0\gamma) = (1.36\pm 0.10)\cdot 10^{-3}$, 
                            
 $ Br(\rho\to\eta^0\gamma) = (2.77\pm 0.31)\cdot 10^{-4}$,             
 $ Br(\omega\to\eta^0\gamma) = (4.22\pm 0.50)\cdot 10^{-4}$. 
	   
   Let us note  that the simple vector meson dominance model
describes  the measured cross section of the $e^+e^-\to \pi^0\gamma$ and $e^+e^-\to             
\eta^0\gamma$ processes very well.

{\bf{Acknowledgments}}  The work is supported in part by grants
Sci.School-1335.2003.2, RFBR 04-02-16181-a, 04-02-16184-a, 05-02-16250-a.

\end{document}